\documentclass{svjour2}                    
\smartqed  
\usepackage[dvipdfmx]{graphicx}
\usepackage{mediabb}

\usepackage{setspace}
\usepackage[usenames]{color} 
\usepackage{amssymb} 
\usepackage{amsmath} 
\usepackage[utf8]{inputenc} 
\usepackage{bm}
\usepackage{cite}
\usepackage{cases}
\newcommand{\ProbP}{\mathbb{P}}

\newcommand{\Vc}[1]{\bm{#1}}
\newcommand{\secref}[1]{Sect.~\ref{#1}}
\newcommand{\fgref}[1]{Fig.~\ref{#1}}
\newcommand{\eqnref}[1]{Eq.~(\ref{#1})}
\newcommand{\Transpose}{\mathbb{T}}
\newcommand{\defeq}{:=}
\newcommand{\average}[1]{\left<#1\right>}
\newcommand{\dd}{\mathrm{d}}
\newcommand{\dt}{\dd t}

\begin{document}

\title{Feedback Regulation and its Efficiency in Biochemical Networks
}
\subtitle{}


\author{ Tetsuya J. Kobayashi \and Ryo Yokota \and Kazuyuki Aihara
}


\institute{Tetsuya J. Kobayashi, Ryo Yokota \& Kazuyuki Aihara \at
              Institute of Industrial Science, the University of Tokyo. 4-6-1 Komaba, Meguro-ku, Tokyo, 153-8505, Japan \\
              Tel.: +81-3-5452-6798\\
              Fax: +81-3-5452-6798\\
              \email{tetsuya@mail.crmind.net}           
}

\date{Received: date / Accepted: date}

\maketitle

\begin{abstract}
Intracellular biochemical networks fluctuate dynamically due to various internal and external sources of fluctuation.
Dissecting the fluctuation into biologically relevant components is important for understanding how a cell controls and harnesses noise and how information is transferred over apparently noisy intracellular networks.
While substantial theoretical and experimental advancement on the decomposition of fluctuation was achieved for feedforward networks without any loop, we still lack a theoretical basis that can consistently extend such advancement to feedback networks. 
The main obstacle that hampers is the circulative propagation of fluctuation by feedback loops. 
In order to define the relevant quantity for the impact of feedback loops for fluctuation, disentanglement of the causally interlocked influence between the components is required.
In addition, we also lack an approach that enables us to infer non-perturbatively the influence of the feedback to fluctuation as the dual reporter system does in the feedforward network.
In this work, we resolve these problems by extending the work on the fluctuation decomposition and the dual reporter system.
For a single-loop feedback network with two components, we define feedback loop gain as the feedback efficiency that is consistent with the fluctuation decomposition for feedforward networks.
Then, we clarify the relation of the feedback efficiency with the fluctuation propagation in an open-looped FF network.
Finally, by extending the dual reporter system, we propose a conjugate feedback and feedforward system for estimating the feedback efficiency only from the statistics of the system non-perturbatively.

\end{abstract}

\section{Introduction}\label{intro}
\subsection{Dissecting fluctuation in biochemical networks}\label{Intro:1}
Biochemical molecules in a cell fluctuate dynamically because of the stochastic nature of intracellular reactions, fluctuation of the environment, and the spontaneous dynamics of intracellular networks\cite{Kaern:2005gra,Raj:2008ipa,Shahrezaei:2008ez}.
Some part of the fluctuation is noise that impairs or disturbs the robust operation of the intracellular networks.
The other part, however, conveys information on complex dynamics of various factors inside and outside of the cell\cite{Kobayashi:2012jia,Purvis:2013dd}.
Dissecting fluctuation into distinct components with different biological roles and meanings is crucial for understanding the mechanisms how a cell controls and harnesses the noise and how information is transferred over apparently noisy intracellular network\cite{Eldar:2010kka,Bowsher:2014cd}.

The intracellular fluctuation is generated from the intracellular reactions or comes from the environment, and then propagates within the intracellular networks. 
The fluctuation of individual molecular species within the networks is therefore a consequence of the propagated fluctuation from the different sources.
Decomposition of the fluctuation into the contributions from the sources is the indispensable step for understanding their biological roles and relevance.
When fluctuation propagates from one component to another unidirectionally without circulation, 
the fluctuation of the downstream can be decomposed into two contributions. 
One is the intrinsic part that originates within the pathway between the components.
The other is the extrinsic part that propagates from the upstream component.
Such decomposition can easily be extended for the network with cascading or branching structures in which no feedback exists. 
This fact drove the intensive anatomical analysis of the intracellular fluctuation in the last decade.

\subsection{Decomposition of sources of fluctuation}
In order to dissect fluctuation into different components, two major strategies have been developed.
One is to use the dependency of each component on different kinetic parameters in the network.
By employing theoretical predictions on such dependency, we can estimate the relative contributions of different components from single-cell experiments with perturbations.
Possible decompositions of the fluctuation were investigated theoretically for various networks such as single gene expression\cite{Ozbudak:2002iqa,Blake:2003cna,Paulsson:2004dha}, signal transduction pathways\cite{Shibata:2005dpa,Ueda:2007iea}, and cascading reactions\cite{Thattai:2002ita}.
Some of them were experimentally tested\cite{Ozbudak:2002iqa,Blake:2003cna,Pedraza:2005jn}.

The other is the dual reporter system in which we simultaneously measure a target molecule with its replica obtained by synthetically duplicating the target.
From the statistics of the target and the replica, i.e, mean, variance, and covariance, we can discriminate the  intrinsic  and extrinsic contributions to the fluctuation because the former is independent between the target and the replica whereas the latter is common to them. 
The idea of this strategy was proposed and developed in \cite{Swain:2002kna,Paulsson:2004dha}, and verified experimentally for different species\cite{Elowitz:2002hba,Raser:2004gha,NeildezNguyen:2007kr}.
Its applicability and generality were further extended \cite{Hilfinger:2011ed,Hilfinger:2012gl,Bowsher:2012kb,Bowsher:2013fn,Rhee:2014hu}.

Now these strategies play the fundamental role to design single-cell experiments and to derive information on the anatomy of fluctuation from the experimental observations\cite{Rosenfeld:2005hnb,Rausenberger:2008kr,Pedraza:2008cp,Taniguchi:2010cba,Cox:2010hi,Hensel:2012cq}.

\subsection{Feedback regulation and its efficiency}\label{Intro:3}
Even with the theoretical and the experimental advancement in decomposing fluctuation, most of works focused on the feedforward (FF) networks in which no feedback and circulation exist.
As commonly known in the control theory\cite{Cosentino:2011up,Savageau:1974wm}, feedback (FB) loops substantially affect fluctuation of a network by either suppressing or amplifying it.
Actually, the suppression of fluctuation in a single gene expression with  a FB loop was experimentally tested in \cite{Becskei:2000fta} earlier than the decomposition of fluctuation. 
While the qualitative and the quantitative impacts of the FB loops were investigated both theoretically and experimentally\cite{Simpson:2003gla,Tomioka:2004bsa,Lestas:2010gq,Oyarzun:2015iz,Austin:2006dqa,Okano:2008gfa,Rhee:2014hu}, we still lack a theoretical basis that can consistently integrate such knowledge with that on the fluctuation decomposition developed for  the FF networks.

The main problem that hampers the integration is the circulation of fluctuation in the FB network.
Because fluctuation generated at a molecular component propagates the network back to itself, 
we need to disentangle the causally interlocked influence between the components to define the relevant quantity for the impact of the FB loops.
From the experimental point of view, in addition, quantification of the impact of FB loops by perturbative experiments  is not perfectly reliable because artificial blocking of the FB loops inevitably accompanies the change not only in fluctuation but also in the average level of the molecular components involved in the loops. 
It is quite demanding and almost impossible for most cases to inhibit the loops by keeping the average level unchanged. 
We still lack an approach that enables us to infer the influence of the FB non-perturbatively as the dual reporter system does.

\subsection{Outline of this work}\label{Intro:4}
In this work, we resolve these problems by extending the work on the fluctuation decomposition\cite{Paulsson:2004dha,Shibata:2005dpa} and the dual reporter system\cite{Swain:2002kna,Elowitz:2002hba}.
By using a single-loop FB network with two components and its linear noise approximation (LNA)\cite{VanKampen:2011vs,Elf:2003hh,Tomioka:2004bsa}, we first provide a definition of the FB loop gain as FB efficiency that is consistent with the fluctuation decomposition in \cite{Paulsson:2004dha,Shibata:2005dpa}.
Then, we clarify the relation of the FB efficiency with the fluctuation propagation in a corresponding open-looped FF network.
Finally, by extending the dual reporter system, we propose a conjugate FB and FF system for estimating the feedback efficiency only from the statistics of the system non-perturbatively.
We also give a fluctuation relation among the statistics that may be used to check the validity of the LNA for a given network.

The rest of this paper is organized as follows.
In \secref{sec:FD}, we review the decomposition of fluctuation for a simple FF system derived in \cite{Paulsson:2004dha,Shibata:2005dpa} by using the LNA.
In \secref{sec:FB}, we extend the result shown in \secref{sec:FD} to a FB network by deriving a decomposition of the fluctuation with feedback.
Using this decomposition, we define the FB loop gain that is relevant for quantifying the impact of the FB  to the fluctuation.
In \secref{sec:FB2}, we give a quantitative relation of the loop gain in the FB network with the fluctuation propagation in a corresponding open-looped FF network.
In \secref{sec:conj}, we propose a conjugate FF and FB network as a natural extension of the dual reporter systems used mainly for the FF networks.
We clarify that the loop gain can be estimated only from the statistics, i.e., mean, variances, and covariances, of the conjugate network.
We also show that a fluctuation relation holds among the statistics, which generalizes the relation used in the dual reporter system. 
In \secref{sec:Dis}, we discuss a link of the conjugate network with the directed information, and give future directions of our work.

\section{Fluctuation Decomposition and Propagation in a Small Biochemical Network}\label{sec:FD}
In this section, we summarize the result for the decomposition of fluctuation obtained in \cite{Paulsson:2004dha,Shibata:2005dpa} by using the LNA, and also its relation with the dual reporter system that was employed in \cite{Elowitz:2002hba,Raser:2004gha,NeildezNguyen:2007kr} to quantify the intrinsic and extrinsic contributions from the experimental measurements.

\subsection{Stochastic chemical reaction and its linear noise approximation}
Let us consider a chemical reaction network consisting of $N$ different molecular species and $M$ different reactions.
We assume that the stochastic dynamics of the network is modeled by the following chemical master equation:
\begin{equation}
\frac{\mathrm{d}\ProbP(t,\Vc{n})}{\mathrm{d}t}=\sum_{k=1}^{M}\left[a_{k}(\Vc{n}-\Vc{s}_{k})\ProbP(t,\Vc{n}-\Vc{s}_{k})- a_{k}(\Vc{n})\ProbP(t,\Vc{n})\right], \label{eq:CME}
\end{equation}
where $\Vc{n}=(n_{1}, \ldots, n_{N})^{\Transpose}\in \mathbb{N}_{\ge 0}^{N}$ is the numbers of the molecular species, $\ProbP(t,\Vc{n})$ is the probability that the number of molecular species is $\Vc{n}$ at $t$, and $a_{k}(\Vc{n}) \in \mathbb{R}_{\ge 0}^{M}$ and $\Vc{s}_{k}$ are the propensity function and the stoichiometric vector of the $k$th reaction, respectively\cite{Gardiner:2009tp,VanKampen:2011vs,gillespie1992rigorous}.
The propensity function characterizes the probability of occurrence of the $k$th reaction when the number of the molecular species is $\Vc{n}$, and the stoichiometric vector defines the change in the number of the molecular species when the $k$th reaction occurs.

In general, it is almost impossible to directly solve \eqnref{eq:CME} both analytically and numerically because it is a high-dimensional or infinite-dimensional differential equation.
To obtain insights for the dynamics of the reaction network, several approximations have been introduced\cite{Gardiner:2009tp,VanKampen:2011vs}. 
Among others, the first-order approximation is the deterministic reaction equation that is described for the given  propensity functions and the stoichiometric vector as 
\begin{equation}
\frac{\mathrm{d}\Omega \Vc{u}(t)}{\mathrm{d}t} = S \Vc{a}(\Omega\Vc{u(t)}), \label{eq:DE}
\end{equation}
where $\Omega$ is the system size, $\Vc{u}\in \mathbb{R}_{\ge 0}^{n}$ is the concentration of $\Vc{n}$ as $\Vc{u}=\Vc{n}/\Omega$, $ \Vc{a}(\Vc{n})\defeq (a_{1}(\Vc{n)}), \ldots, a_{M}(\Vc{n}))^{\Transpose}$, and $S_{*,k}\defeq\Vc{s}_{k}$ is the stoichiometric matrix.
Equation (\ref{eq:DE}) was successfully applied for chemical reaction networks with large system size where the fluctuation of the concentration of the molecular species can be neglected. 
When the system size is not sufficiently large, however, \eqnref{eq:DE} is not relevant for analyzing the fluctuation of the network.
The LNA is a kind of the second-order approximation of the \eqnref{eq:CME} that characterizes the fluctuation around a fixed point, $\bar{\Vc{n}}$, of \eqnref{eq:DE} that satisfies $ S \Vc{a}(\Vc{\bar{n}})=0$.
The stationary fluctuation of the network around $\bar{\Vc{n}}$ is then obtained by solving the following Lyapunov equation\cite{VanKampen:2011vs,Elf:2003hh,Tomioka:2004bsa}
\begin{equation}
(K(\bar{\Vc{n}})\Vc{\Sigma})+(K(\bar{\Vc{n}})\Vc{\Sigma})^{\Transpose}+D(\bar{\Vc{n}})=0, \label{eq:LE}
\end{equation}
where 
\begin{equation}
K_{*,j}(\Vc{n})\defeq\frac{\partial S\Vc{a}(\Vc{n})}{\partial n_{j}},\qquad D_{i,j}(\Vc{n})\defeq\sum_{k}  s_{i,k}s_{j,k}a_{k}(\Vc{n}),
\end{equation}
and $\Vc{\Sigma}$ is the covariance matrix of $\Vc{n}$.
When the propensity function $\Vc{a}(\Vc{n})$ is affine with respect to $\Vc{n}$,  the dynamics of $\Omega \Vc{u}(t)$ determined by \eqnref{eq:DE} is identical to that of the first cumulant of $n$, i.e., the average of $n$, as 
\begin{equation}
\frac{\mathrm{d}\left<\Vc{n}\right>}{\mathrm{d}t} = S \Vc{a}(\left<\Vc{n}\right>),\label{eq:av}
\end{equation}
where $\left<\Vc{n}\right> \defeq \sum_{\Vc{n}}\Vc{n}\ProbP(t,\Vc{n})$. 
In addition, the second cumulant, i.e., the covariance matrix, follows the Lyapunov equation as 
\begin{equation}
\frac{\mathrm{d} \Vc{\Sigma}}{\mathrm{d}t}=(K(\left<\Vc{n}\right> )\Vc{\Sigma})+(K(\left<\Vc{n}\right> )\Vc{\Sigma})^{\Transpose}+D(\left<\Vc{n}\right> ).\label{eq:LE2}
\end{equation}
Therefore, if  the propensity function $\Vc{a}(\Vc{n})$ is affine, the stationary fluctuation of $\Vc{n}$ is exactly described by \eqnref{eq:LE}.
For a non-affine $\Vc{a}(\Vc{n})$, \eqnref{eq:av} and \eqnref{eq:LE2} can also be regarded as an approximation of the full cumulant equations by the cumulant closure\cite{Gardiner:2009tp} under which we ignore the influence of the second and the higher order cumulants to \eqnref{eq:av}, and that of the third and the higher order ones to \eqnref{eq:LE2}\footnote{We prefer this interpretation of the Lyapunov equation because the LNA has been applied for various intracellular networks whose system size is not sufficiently large.}.
Even though the propensity function $\Vc{a}(\Vc{n})$ is not affine,  \eqnref{eq:LE} (or \eqnref{eq:LE2}) can produce a good approximation of the fluctuation, provided that the fixed point, $\bar{\Vc{n}}$, is a good approximation of the average, $\left<\Vc{n} \right>$, and that the local dynamics around the fixed point is approximated enough by its linearization.
In addition, compared with other approximations, the LNA enables us to obtain an analytic representation of the fluctuation because \eqnref{eq:LE} is a linear algebraic equation with respect to $\Vc{\Sigma}$.
Owing to this property, the LNA and its variations played the crucial role to reveal the analytic representation of the fluctuation decomposition in biochemical reaction networks\cite{Tomioka:2004bsa,Paulsson:2004dha}. 
As in these previous works, we employ the LNA to obtain an analytic representation for the feedback efficiency.

\begin{figure}
\includegraphics[width=\textwidth]{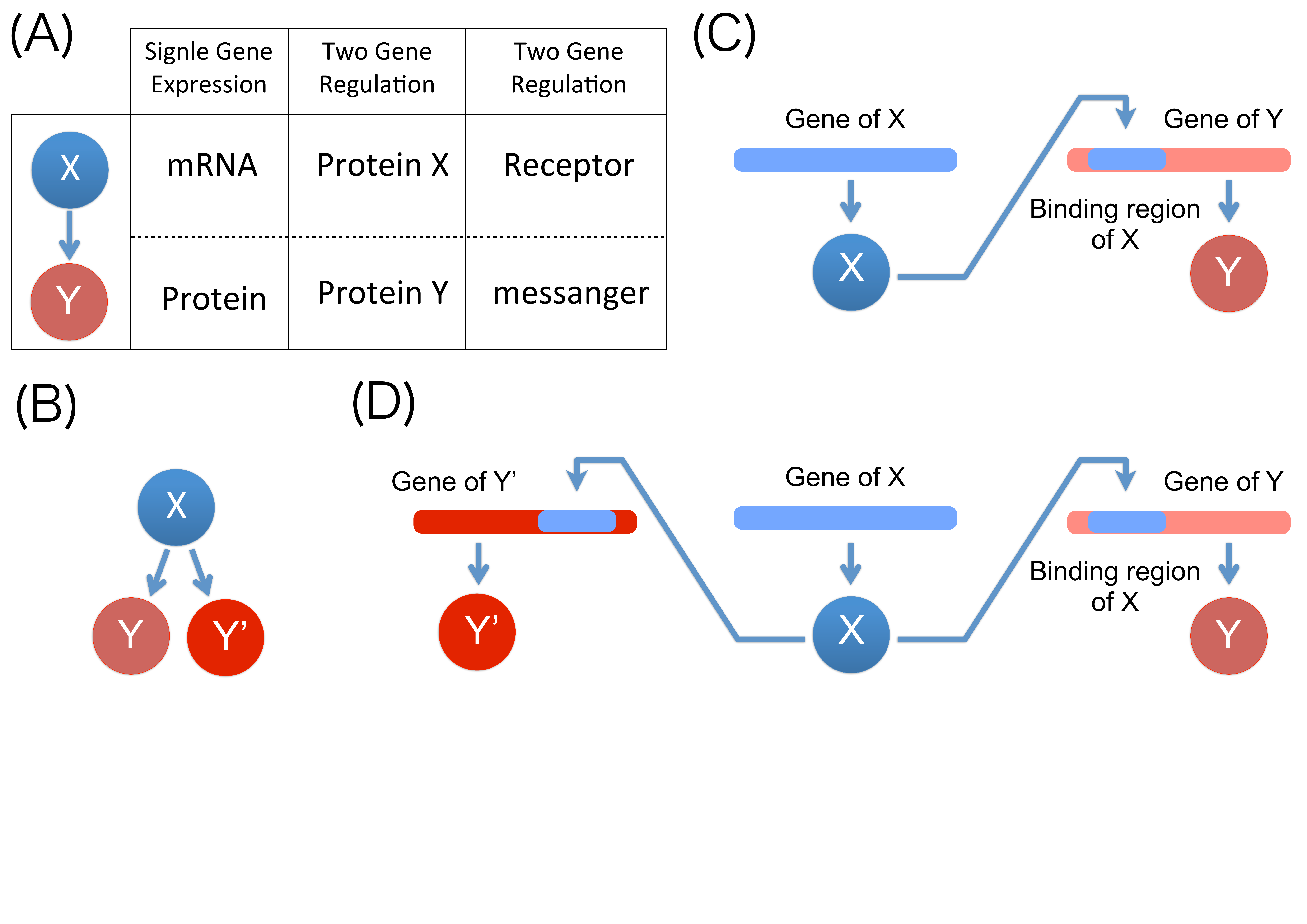}
\vspace{-2.3cm}
\caption{(A) The structure of the two-component FF network. Interpretations of this network as single-gene expression\cite{Ozbudak:2002iqa,Elowitz:2002hba,Paulsson:2004dha,Morishita:2004iva}, two-gene regulation\cite{Tomioka:2004bsa}, and signal transduction \cite{Ueda:2007iea} are shown.  (B) The structure of the dual reporter system\cite{Swain:2002kna,Elowitz:2002hba,Paulsson:2004dha,Hilfinger:2011ed,Bowsher:2012kb}. 
(C) A schematic diagram of the FF network for two-gene regulation.(D) A schematic diagram of the dual reporter network for the two-gene regulation.}
\label{fig1}       
\end{figure}

\subsection{Decomposition of fluctuation}
Starting from the LNA, Paulsson derived an analytic result on how noise is determined in a FF network with two components (\fgref{fig1} (A))\cite{Paulsson:2004dha}.
Here, we briefly summarize the result derived in  \cite{Paulsson:2004dha}.
Let $n_{1}=x$ and $n_{2}=y$ for notational simplicity, and consider the FF reaction network (\fgref{fig1} (A)) with the following propensity function and stoichiometric matrix,
\begin{align}
\Vc{a}(x,y)=(a_{x}^{+}(x),a_{x}^{-}(x),a_{y}^{+}(x,y),a_{y}^{-}(x,y)),\qquad 
S=\begin{pmatrix}
+1 & -1 & 0 & 0 \\
0   &   0 & +1 & -1
\end{pmatrix}.
\end{align}
Because $a_{x}^{\pm}$ depends only on $x$, $x$ regulates $y$ unidirectionally.
Then, for a fixed point $(\bar{x},\bar{y})$ of \eqnref{eq:DE}, that satisfies
\begin{align}
a_{x}^{+}(\bar{x}) =a_{x}^{-}(\bar{x}), \quad  a_{y}^{+}(\bar{x},\bar{y})=a_{y}^{-}(\bar{x},\bar{y}),
\end{align}
$K$ and $D$ in \eqnref{eq:LE} becomes
\begin{align}
K&=\begin{pmatrix}
-d_{x} & 0\\
k_{yx}& -d_{y}
\end{pmatrix}=
\begin{pmatrix}
-H_{xx}/\tau_{x} & 0\\
-\frac{\bar{y}}{\bar{x}}H_{yx}/\tau_{y}& -H_{yy}/\tau_{y}
\end{pmatrix}, \quad 
D=\begin{pmatrix}
 2 \bar{a}_{x} & 0 \\
0 &  2 \bar{a}_{y}
\end{pmatrix},\label{eq:KDFF}
\end{align}
where $\bar{a}_{x} \defeq a_{x}^{+}(\bar{x}) =a_{x}^{-}(\bar{x})$, $\bar{a}_{y}=a_{y}^{+}(\bar{x},\bar{y})=a_{y}^{-}(\bar{x},\bar{y})$, and $H_{i,j}\defeq\left.\frac{\ln a_{i}^{-}/a_{i}^{+}}{\partial \ln j}\right|_{(\bar{x},\bar{y})}$ for $i,j \in \{x,y\}$.
$d_{x}$ and $d_{y}$ are the minus of the diagonal terms of $K$ and represent the effective degradation rates of $x$ and $y$. 
$k_{yx}$ is the off-diagonal term of $K$ that represents the interaction from $x$ to $y$. 

$H_{ij}$ is the susceptibility of the $i$th component to the perturbation of the $j$th one. 
$\tau_{x}\defeq \bar{x}/\bar{a}_{x}$ and $\tau_{y}\defeq \bar{y}/\bar{a}_{y}$ are the effective life-time of $x$ and $y$, respectively.
Except this section, we mainly use $d$s and $k$s as the representation of the parameters rather than $H$s and $\tau$s introduced in \cite{Paulsson:2004dha}\footnote{The former gives a notationally simpler result.}.
By solving \eqnref{eq:LE} analytically,  the following fluctuation-dissipation relation was derived in \cite{Paulsson:2004dha} as
\begin{align}
\frac{\sigma_{x}^{2}}{\bar{x}^{2}}=\underbrace{\frac{1}{\bar{x} H_{xx}}}_{(I)},\quad 
\frac{\sigma_{y}^{2}}{\bar{y}^{2}}=\underbrace{\frac{1}{\bar{y} H_{yy}}}_{(II)} + \underbrace{\overbrace{\left(\frac{H_{yx}/\tau_{y}}{H_{yy}/\tau_{y}}\right)^{2}}^{(iii)}\overbrace{\frac{H_{yy}/\tau_{y}}{H_{yy}/\tau_{y}+H_{xx}/\tau_{x}}}^{(ii)}\overbrace{\frac{\sigma_{x}^{2}}{\bar{x}^{2}}}^{(i)}}_{(III)}\label{eq:FD}.
\end{align}
This representation measures the intensity of the fluctuation by the coefficient of variation (CV)\footnote{The CV is defined by the ratio of the standard deviation to the mean as $\sigma_{x}/\average{x}$.}, and describes how the fluctuation generates and propagates within the network. 
$(I)$ is the intrinsic fluctuation of $x$ that originates from the stochastic birth and death of $x$.
$1/\bar{x}$ reflects the Poissonian nature of the stochastic birth and death, and $1/H_{xx}$ is the effect of auto-regulatory FB.
Similarly, $(II)$ is the intrinsic fluctuation of $y$ that originates from the stochastic birth and death of $y$.
$(III)$, on the other hand, corresponds to the extrinsic contributions to the fluctuation of $y$ due to the fluctuation of $x$.
The term $(III)$ is further decomposed into $(i)$, $(ii)$, and $(iii)$. 
$(i)$ is the fluctuation of $x$, and therefore, identical to $(I)$.
$(ii)$ and $(iii)$ determine the efficiency of the propagation of the fluctuation from $x$ to $y$, which
correspond to the time-averaging and sensitivity of the pathway from $x$ to $y$, respectively.
This representation captures the important difference of the intrinsic and the extrinsic fluctation such that the intrinsic one, the term $(II)$,  can be reduced by increasing the average of $\bar{y}$ whereas the extrinsic one, the term $(III)$,  cannot.  

While \eqnref{eq:FD} provides an useful interpretation on how the fluctuation propagates in the FF network, 
it is not appropriate for the extension to the FB network because the contribution of $x$ to the fluctuation of $y$ is described by the CV of $x$ as the term $(i)$.
Because the fluctuation of $x$ and $y$ depend mutually if we have a FB between $x$ and $y$, we need to characterize the fluctuation of $y$ without directly using the fluctuation of $x$. 
To this end, we adopt the variances  and covariances as the measure of the fluctuation  and use the following decomposition of the fluctuation for the FF network:
\begin{align}
\sigma_{x}^{2}=\underbrace{\overbrace{G_{x,x}}^{(ii)}\overbrace{\bar{x}}^{(i)}}_{(I)},\qquad 
\sigma_{y}^{2}=\underbrace{\overbrace{G_{yy}}^{(ii)}\overbrace{\bar{y}}^{(i)}}_{(II)} + \underbrace{\overbrace{G_{yx}}^{(iii)}\overbrace{G_{x,x}}^{(ii)}\overbrace{\bar{x}}^{(i)}}_{(III)},\label{eq:FD2}
\end{align}
where we define $G_{xx}$, $G_{yy}$, and $G_{yx}$ as 
\begin{align}
G_{xx}=\frac{1}{H_{xx}},\quad G_{yy}=\frac{1}{H_{yy}}, \quad G_{yx}=  \frac{k_{yx}^{2}}{(d_{x}+d_{y})d_{y}}.
\end{align}
The terms $(I)$, $(II)$, and $(III)$ in \eqnref{eq:FD2} correspond to those in \eqnref{eq:FD}.
The interpretation of the terms within $(I)$, $(II)$, and $(III)$ is, however, different.
In \eqnref{eq:FD2}, the terms $(i)$ are interpreted as the fluctuation purely generated by the birth and death reactions of $x$ and $y$ by neglecting any contribution of the auto-FBs.
Because the simple birth and death of a molecular species without any regulation follow the Poissonian statistics, the intensity of the fluctuation is equal to the means of the species.
Thus, $\bar{x}\approx \average{x}$ and $\bar{y}\approx \average{y}$ in the terms $(i)$ represent the generation of the fluctuation by the birth and death of $x$ and $y$, respectively.
The fluctuation generated is then amplified or suppressed by the auto-regulatory FBs.
$G_{xx}$ and $G_{yy}$ in the terms $(ii)$ account for this influence, and are denoted as auto-FB gains in this work.
Finally, $G_{yx}$ in the term $(iii)$ quantifies the efficiency of the propagation of the fluctuation from $x$ to $y$. 
We denote $G_{yx}$ as the path gain from $x$ to $y$.
If we use the notation in \eqnref{eq:FD}, $G_{yx}$ is described as 
\begin{align}
G_{yx}=  \left(\frac{\bar{y}H_{yx}/\tau_{y}}{\bar{x}H_{yy}/\tau_{y}}\right)^{2}\frac{H_{yy}/\tau_{y}}{H_{yy}/\tau_{y}+H_{xx}/\tau_{x}}.
\end{align}

The decomposition of the fluctuation of $y$ into $(II)$ and $(III)$ is consistent between \eqnref{eq:FD} and \eqnref{eq:FD2} while the further decompositions within $(II)$ and $(III)$ are different.
It was recently revealed in \cite{Hilfinger:2011ed,Bowsher:2012kb,Hilfinger:2012gl,Bowsher:2013fn} that the decomposition into $(II)$ and $(III)$ is linked to the variance decomposition formula in statistics as 
\begin{align}
\sigma_{y}^{2}&=\underbrace{\mathbb{V}[\mathbb{E}[y(t)|\mathcal{X}(t)]]}_{(II)} + \underbrace{\mathbb{E}[\mathbb{V}[y(t)|\mathcal{X}(t)]]}_{(III)}, \label{eq:VDF}
\end{align}
where $\mathcal{X}(t)\defeq\{x(\tau);\tau \in [0,t]\}$ is the history of $x(t)$, and $\mathbb{E}$ and $\mathbb{V}$ are the expectation and the variance, respectively\footnote{The correspondence of \eqnref{eq:VDF} with \eqnref{eq:FD} or \eqnref{eq:FD2} is valid only when $\sigma_{y}^{2}$ is decomposed under the conditioning with respect to the history of $x(t)$, $\mathcal{X}(t)$, rather than the instantaneous state of $x(t)$ at $t$. }.
Because the variance decomposition formula holds generally, \eqnref{eq:VDF} is more fundamental than \eqnref{eq:FD} and \eqnref{eq:FD2} as the decomposition of fluctuation. 
This decomposition was further analyzed in \cite{Bowsher:2012kb,Hilfinger:2012gl,Bowsher:2013fn}.

\subsection{Dual reporter system}
The decomposition, \eqnref{eq:FD} or \eqnref{eq:FD2},  guides us how to evaluate the intrinsic and the extrinsic fluctuation in $y$ experimentally. 
When we can externally control the mean of $y$ without affecting the term $(III)$, we can estimate the relative contributions of $(II)$ and $(III)$ as the intrinsic and the extrinsic fluctuation by  plotting $\sigma_{y}^{2}$ as a function of the mean of $y$.
When $x$ and $y$ correspond to a mRNA and a protein in the single gene expression, the translation rate works as such a control parameter\cite{Morishita:2004iva}.
This approach was intensively employed to estimate the efficiency of the fluctuation propagation in various intracellular networks\cite{Ozbudak:2002iqa,Blake:2003cna,Pedraza:2005jn}.

Another way to quantify the intrinsic and the extrinsic fluctuation is the dual reporter system adopted in \cite{Elowitz:2002hba,Raser:2004gha,NeildezNguyen:2007kr} whose network structure is shown in \fgref{fig1} (B)\footnote{The dual reporter system is also called a conjugate reporter system\cite{Bowsher:2012kb,Bowsher:2013fn}.}. 
In the dual reporter system, a replica of $y$ is attached to the downstream of $x$ as in \fgref{fig1} (B) where $y'$ denotes the molecular species of the replica.
The replica, $y'$, must have the same kinetics as $y$, and must be measured simultaneously with $y$. 
If $y$ is a protein whose expression is regulated by another protein, $x$, as in \fgref{fig1} (C),  then $y'$ can be synthetically constructed by duplicating the gene of $y$ and attaching fluorescent probes with different colors to $y$ and $y'$ as in \fgref{fig1} (D) \cite{Elowitz:2002hba}.
Under the LNA, the covariance between $y$ and $y'$ can be described  as 
\begin{align}
\sigma_{y,y'}=G_{yx}G_{xx}\bar{x}.
\end{align}
Thus, by using only the statistics of the dual reporter system, the intrinsic and the extrinsic components in $y$ can be estimated as 
\begin{align}
\sigma_{y}^{2}=\underbrace{G_{yy}\average{y}}_{(II)} +  \underbrace{\sigma_{y, y'}}_{(III)},\label{eq:DRSstat}
\end{align}
where it is unnecessary to control  any kinetic parameters externally.
This approach is generalized as the conjugate reporter system together with the more general decomposition, \eqnref{eq:VDF}, in \cite{Bowsher:2012kb,Hilfinger:2012gl,Bowsher:2013fn}.

\section{Feedback Loop Gain in a Small Biochemical Network}\label{sec:FB}
\begin{figure}
\includegraphics[width=\textwidth]{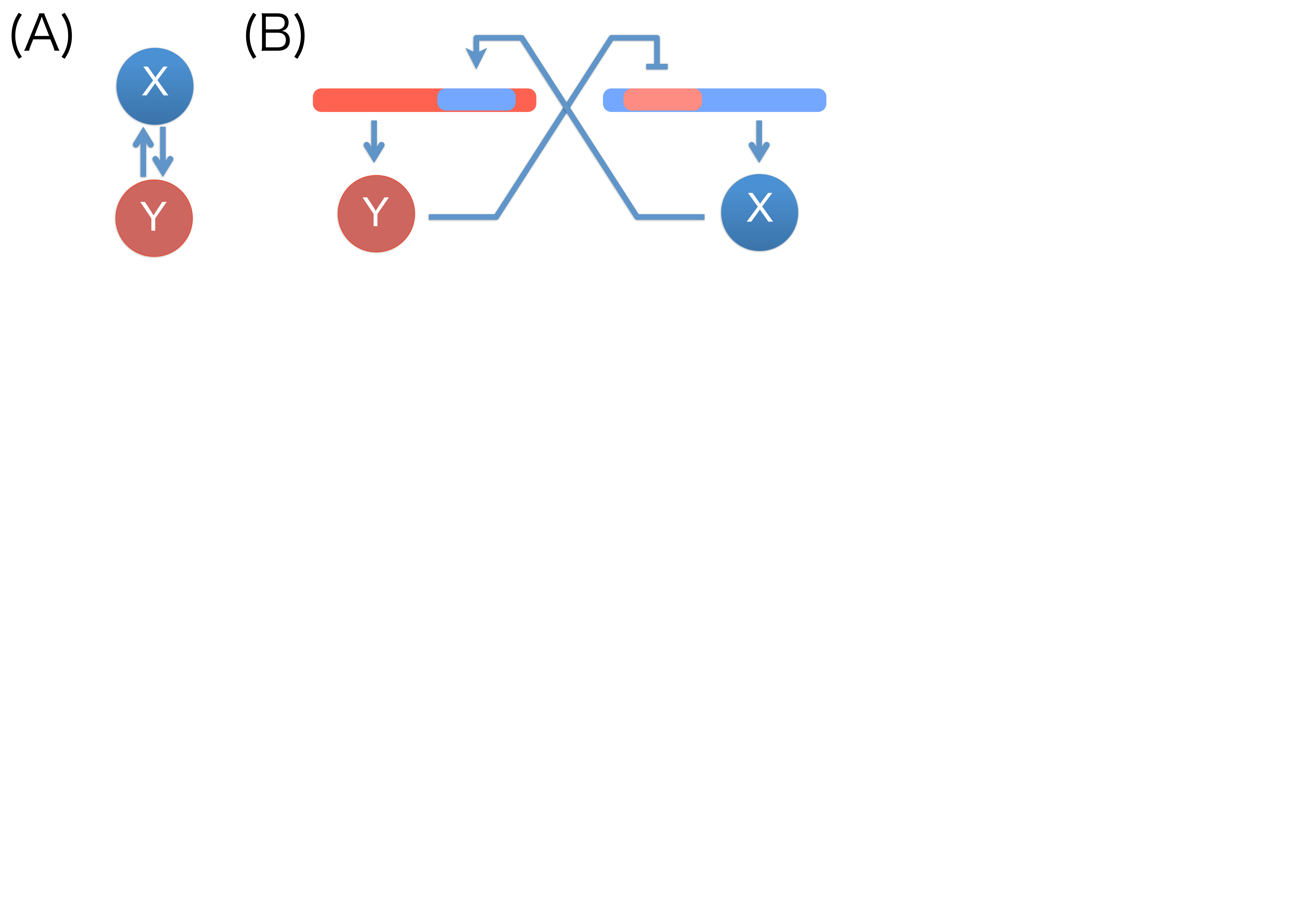}
\vspace{-6cm}
\caption{(A) The structure of the two-component FB network. (B) A schematic diagram of the FB network for two-gene regulation. }
\label{fig2}       
\end{figure}

While we had substantial progress in the decomposition of the fluctuation and its experimental measurement for the FF networks in the last decade, its extension to FB networks is yet to be achieved.
In this section, we extend the decomposition of the fluctuation in the simple FF network (\eqnref{eq:FD2}) to the corresponding FB one depicted in \fgref{fig2} (A) .  
As in \eqnref{eq:KDFF}, $K$ and $D$ in the Lyapunov equation (\eqnref{eq:LE}) for the FB network can be described as
\begin{align}
K=\begin{pmatrix}
-d_{x} & k_{xy} \\
k_{yx} & -d_{y}
\end{pmatrix}
=\begin{pmatrix}
-H_{xx}/\tau_{x} & -\frac{\bar{x}}{\bar{y}}H_{xy}/\tau_{x}\\
-\frac{\bar{y}}{\bar{x}}H_{yx}/\tau_{y}& -H_{yy}/\tau_{y}
\end{pmatrix}
,\qquad D=\begin{pmatrix}
2 \bar{a}_{x} & 0 \\
0 & 2 \bar{a}_{y}.
\end{pmatrix}
\end{align}
By defining the path gain from $y$ back to $x$ as  
\begin{align}
G_{xy}\defeq\frac{k_{xy}^{2}}{d_{x}(d_{x}+d_{y})},
\end{align}
we can derive a decomposition of the fluctuation of $x$ and $y$ as
\begin{align}
\begin{split}
\sigma_{x}^{2}&=\frac{1-L_{y}}{1-L_{x}-L_{y}}\overbrace{G_{xx}\bar{x}}^{(I)} + \frac{1}{1-L_{x}-L_{y}}\overbrace{G_{xy}G_{yy}\bar{y}}^{(IV)},\\
\sigma_{y}^{2}&=\frac{1-L_{x}}{1-L_{x}-L_{y}}\underbrace{G_{yy}\bar{y}}_{(II)} + \frac{1}{1-L_{x}-L_{y}}\underbrace{G_{yx}G_{xx}\bar{x}}_{(III)},
\end{split}\label{eq:FDFB}
\end{align}
where we define
\begin{align}
L_{x}\defeq\frac{k_{xy}k_{yx}}{d_{x}(d_{x}+d_{y})},\qquad L_{y}\defeq\frac{k_{xy}k_{yx}}{(d_{x}+d_{y})d_{y}}\label{eq:LG}.
\end{align}
If the FB from $y$ to $x$ does not exist, i.e., $k_{xy}=0$, then $L_{x}=L_{y}=G_{xy}=0$ and \eqnref{eq:FDFB} is reduced to \eqnref{eq:FD2}.
Thus, $L_{x}$ and $L_{y}$ account for the effect of the FB.
$L_{x}$ and $L_{y}$ are denoted as the FB loop gains in this work.
\eqnref{eq:FDFB} clearly demonstrates that the representation in \eqnref{eq:FD} does not work with the FB because $\sigma_{y}^{2}$ cannot be described with $\sigma_{x}^{2}$ any longer. 
In contrast, we can interpret the terms in \eqnref{eq:FDFB}  consistently with those in \eqnref{eq:FD2} because all the terms, $(I)$, $(II)$ and $(III)$, are unchanged in \eqnref{eq:FDFB}. 
The new term, $(IV)$, in the expression for $\sigma_{x}^{2}$ appears to account for the propagation of the fluctuation generated by the birth and death events of $y$ back to $x$.

Equations (\ref{eq:FDFB}) and (\ref{eq:LG}) indicate how the FB affects the fluctuation of $x$ and $y$. 
First, the fluctuation is suppressed when $L_{x}$ and $L_{y}$ are negative whereas it is amplified when they are positive.
When $d_{x}$ and $d_{y}$ are positive\footnote{For biologically relevant situations, $d_{x}$ and $d_{y}$ are positive because they can be regarded as the effective degradation rates.}, the sign of $L_{x}$ and $L_{y}$ are determined by the sign of $k_{xy}k_{yx}$. 
Thus, the FB loop is negative when $x$ regulates $y$ positively and $y$ does $x$ negatively or vise versa.
This is consistent with the normal definition of the sign of a FB loop.
Second, the efficiency of the FB depends on the source of the fluctuation.
For example, when $L_{x}\ll L_{y}$, e.g., the time-scale of $x$ is much faster than $y$, then \eqnref{eq:FDFB} can be approximated as
\begin{align*}
\sigma_{x}^{2}&=\overbrace{G_{xx}\bar{x}}^{(I)} + \frac{1}{1-L_{y}}\overbrace{G_{xy}G_{yy}\bar{y}}^{(IV)},\quad
\sigma_{y}^{2}=\frac{1}{1-L_{y}}\left[ \underbrace{G_{yy}\bar{y}}_{(II)} + \underbrace{G_{yx}G_{xx}\bar{x}}_{(III)}\right].
\end{align*}
Thus, the FB does not work for the term $(I)$ that is the part of fluctuation of $x$ whose origin is the birth and death events of $x$ itself. 
This result reflects the fact that the slow FB from $x$ to itself via $y$ cannot affect the fast component of the fluctuation of $x$\footnote{Note that the term $(III)$ is  affected by the FB even though its origin is the fast birth and death events of $x$. This can be explained as follows.  In general, the path gain from $x$ to $y$, $G_{yx}$, becomes very small compared to others when $x$ has much faster time scale than $y$. Thus, the term $(III)$ becomes quite small and little fluctuation propagates from $x$ to $y$ because of the averaging effect of the slow dynamics of $y$. $(III)$ represents, therefore, the slow component in the fluctuation of the birth and death of $x$ that has the comparative timescale as that of $y$. This is why the slow FB can affect the term $(III)$.}.
Finally, when $L_{x}$ and $L_{y}$ satisfy $1-L_{x}-L_{y}=1-k_{xy}k_{yx}/d_{x}d_{y}=0$, the fluctuation of both $x$ and $y$ diverges due to the FB.
Since this condition means that the determinant of $K$ becomes $0$ and $K$ is the Jacobian matrix of \eqnref{eq:DE}, 
the fluctuation of $x$ and $y$ diverges due to the destabilization of the fixed point, $\bar{\Vc{n}}$, by the FB.

\section{Relation between Fluctuation Propagation and Feedback Gain}\label{sec:FB2}
As shown in \secref{sec:FB}, $L_{x}$ and $L_{y}$ are quantitatively related to the efficiency of the FB.
Because  $L_{x}L_{y}=G_{xy}G_{yx}$ holds, the loop gains are also linked to the propagation of the fluctuation from $x$ to $y$ and from $y$ back to $x$.
However, the meaning of the individual gains, $L_{x}$ and $L_{y}$, is still ambiguous.

\begin{figure}
\includegraphics[width=\textwidth]{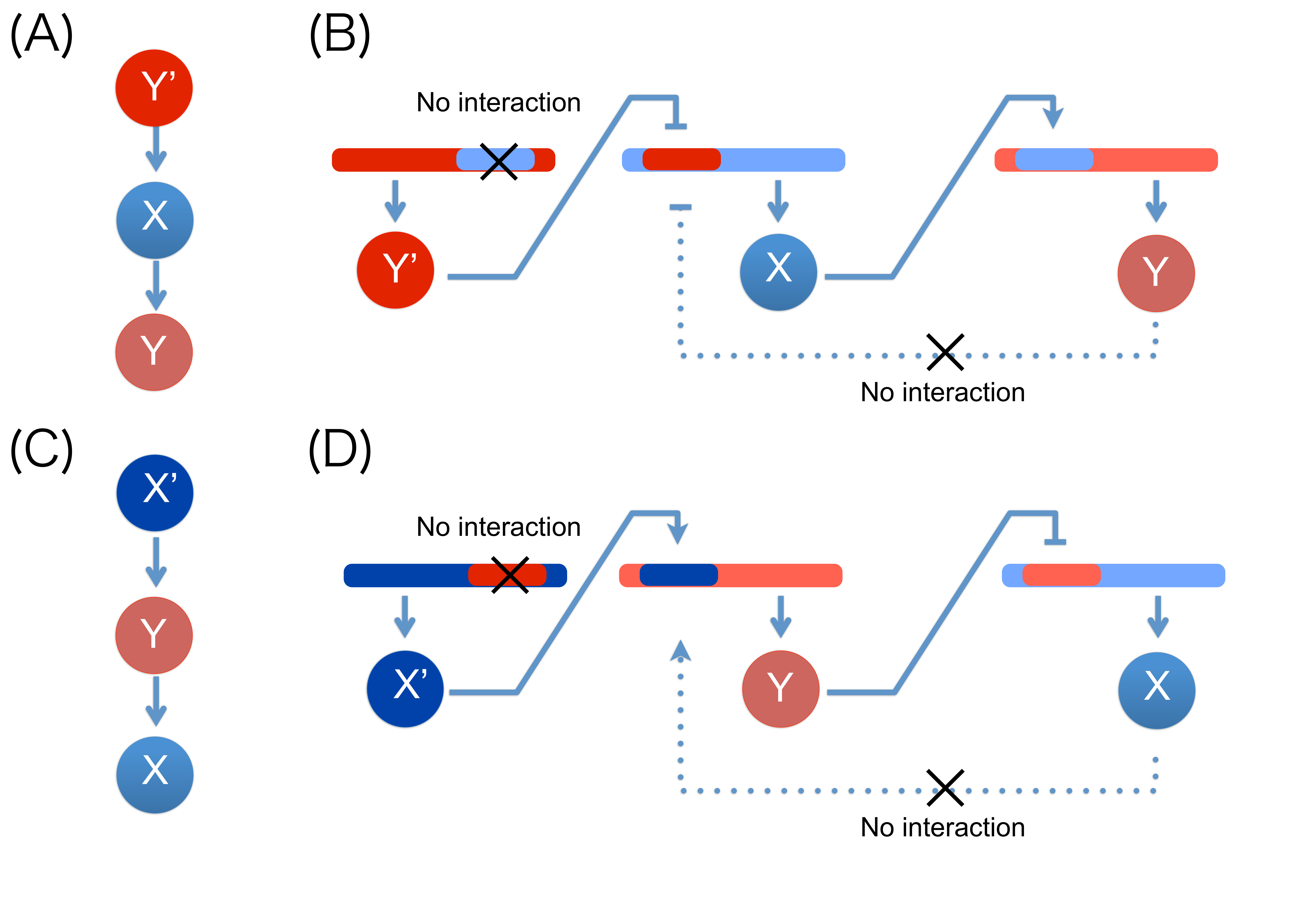}
\vspace{-1cm}
\caption{(A) The structure of the the opened FF network obtained by replicating $y$ in the FB network. (B) A schematic diagram of the network in (A) for gene regulation. (C) The structure of the the opened FF network obtained by replicating $x$ in the FB network. (D) A schematic diagram of the network in (C) for gene regulation.}
\label{fig3}       
\end{figure}

To clarify the relation between the loop gains and the propagation of fluctuation, we consider a three-component FF network shown in \fgref{fig3} (A) and (B) that are obtained by opening the FB network in \fgref{fig2} (A). 
In the network shown in \fgref{fig3} (A), $x$ is regulated not by $y$ but by its replica denoted as $y'$.
We assume that $y'$ is not driven by $x$, and thereby, $y' \to x \to y$ forms a FF network. 
$K$ and $D$ in \eqnref{eq:LE} for this network become
\begin{align}
K=\begin{pmatrix}
-d_{y'} & 0 & 0 \\
k_{y'x} & -d_{x} & 0 \\
0 & k_{xy} & - d_{y}
\end{pmatrix},\qquad 
D=\mathrm{diag}\begin{pmatrix}
\bar{a}_{y'} &  \bar{a}_{x} & \bar{a}_{y} 
\end{pmatrix}.
\end{align}
Because $y'$ is the replica of $y$, we assume that all the kinetic parameters of $y'$ are equal to those of $y$ as $k_{y'x}=k_{yx}$, $d_{y'}=d_{y}$, and $\bar{a}_{y'}=\bar{a}_{y}$.

By solving \eqnref{eq:LE}, we can obtain the following decomposition of the fluctuation of $y$ as
\begin{align}
\begin{split}
\sigma_{x}^{2}&=\overbrace{G_{xx}\bar{x}}^{(I)} + \overbrace{G_{xy'}G_{y'y'}\bar{y}'}^{(IV)},\\
\sigma_{y}^{2}&=\underbrace{G_{yy}\bar{y}}_{(II)} + \underbrace{G_{yx}G_{xx}\bar{x}}_{(III)} + \underbrace{\left[G_{yy'}+G_{yxy'}\right]G_{y'y'}\bar{y}'}_{(V)},
\end{split} \label{eq:FDFF3}
\end{align}
where 
\begin{align}
G_{xy'}=G_{xy},\quad G_{yxy'} &\defeq G_{yx}G_{xy'}=G_{yx}G_{xy}=L_{x}L_{y},\quad G_{yy'}\defeq L_{y}^{2}/2.
\end{align}
The term $(IV)$ for $x$ is similar to the propagation of the fluctuation from $y$ to $x$ in the FB network, but it represents the propagation of the fluctuation from the replica $y'$ in this opened FF network.
The new term $(V)$ accounts for the propagation of the fluctuation from $y'$ down to $y$.
The gain of this propagation has two terms, $G_{yxy'}$ and $G_{yy'}$. 
The first term, $G_{yxy'}$, is the total gain of the fluctuation propagation from $y'$ to $x$ and from $x$ to $y'$ because $G_{yxy'}=G_{yx}G_{xy'}$ holds.
In order to see the meaning of the second term $G_{yy'}$, we need to rearrange \eqnref{eq:FDFF3} as
\begin{align}
\sigma_{y}^{2}&=G_{yy}\bar{y} + G_{yx}\sigma_{x}^{2} + G_{yy'}G_{y'y'}\bar{y}'.
\end{align}
This representation clarifies that $G_{yy'}$ describes the propagation of the fluctuation from $y'$ to $y$ that cannot be reflected to the fluctuation of the intermediate component, $x$.
In addition, by solving \eqnref{eq:LE}, we can see that the gain $G_{yy'}$ is directly related to the covariance between $y$ and $y'$ as
\begin{align}
\sigma_{y',y}=\sqrt{G_{yy'}}\bar{y'}=\frac{L_{y}}{2}\bar{y'}.\label{eq:FFcovyy}
\end{align}
This implies that we have at least two types of propagation of the fluctuation.
One described by $G_{yxy'}$ is that the fluctuation of the upstream, i.e., $y'$, is absorbed by the intermediate component, i.e., $x$, and then the absorbed fluctuation propagates to the downstream, i.e., $y$. 
This component does not convey the information of the upstream because that does not affect the covariance between the upstream and the downstream.
The other described by $G_{yy'}$ is that the fluctuation of the upstream propagates to the downstream without affecting the intermediate component. 
This fluctuation conveys the information on the upstream to the downstream because it is directly linked to their covariance.
The fact that $L_{y}$ is related to the latter indicates that the FB efficiency is directly linked to the information transfer in the opened loop from $y'$ to $y$.
By considering another opened loop network where the replica of $x$ is introduced as in \fgref{fig3} (C), we can obtain the following result:
\begin{align}
\begin{split}
\sigma_{y}^{2}&=\overbrace{G_{yy}\bar{y}}^{(II)} + \overbrace{G_{yx'}G_{x'x'}\bar{x}'}^{(III)},\\
\sigma_{x}^{2}&=\underbrace{G_{xx}\bar{x}}_{(I)} + \underbrace{G_{xy}G_{yy}\bar{y}}_{(IV)} + \underbrace{\left[G_{xx'}+G_{xyx'}\right]G_{x'x'}\bar{x}'}_{(VI)},
\end{split}
 \label{eq:FDFF3x}
\end{align}
where $G_{xx'}\defeq L_{x}^{2}/2$, and we also have
\begin{align}
\sigma_{x',x}&=\sqrt{G_{xx'}}\bar{x'}=\frac{L_{x}}{2}\bar{x'}. \label{eq:FFcovxx}
\end{align}
By combining \eqnref{eq:FFcovyy} and \eqnref{eq:FFcovxx}, we can estimate the loop gains, $L_{x}$ and $L_{y}$, by only measuring the averages and covariances of the opened networks as follows:
\begin{align}
\frac{L_{x}}{2}=\frac{\sigma_{x',x}}{\average{x'}},\quad \frac{L_{y}}{2}=\frac{\sigma_{y',y}}{\average{y'}}. \label{eq:LsOpen}
\end{align}
While this strategy sounds eligible theoretically, it has an experimental difficulty in constructing the opened networks.
In the opened network, the replica, e.g., $y'$, must be designed so that it is free from the regulation of $x$ by keeping all the other properties and kinetic parameters the same as those of $y$.
For example, if $x$ and $y$ are regulatory proteins and if they regulate each other as transcription factors as \fgref{fig2} (B), 
the replica, $y'$, must not be regulated by $x$ but its expression rate must be  equal to the average expression rate of $y$ under the regulation of $x$ as in \fgref{fig3} (B).
This requires fine-tuning of the expression rate of $y'$ by modifying the DNA sequences relevant for the rate.
In order to conduct this tuning, we have to measure several kinetic parameters of the original FB networks that undermines the advantage of the opened network that measurements of the kinetic parameters are unnecessary to estimate $L_{x}$ and $L_{y}$ via \eqnref{eq:LsOpen}.

\section{Estimation of Feedback Loop Gain by a Conjugate FB \& FF Network}\label{sec:conj}
\begin{figure}
\includegraphics[width=\textwidth]{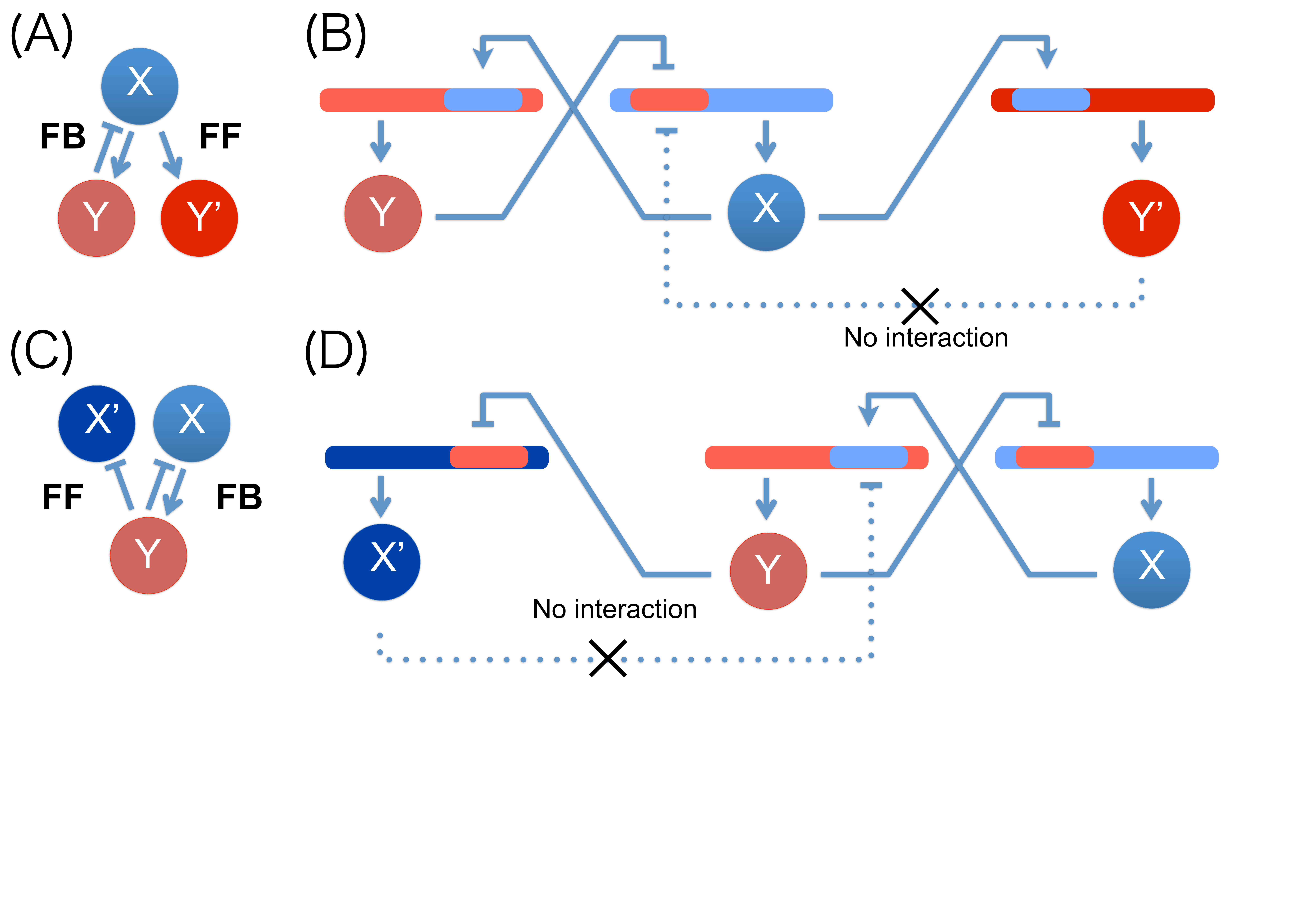}
\vspace{-2.3cm}
\caption{(A) The structure of the conjugate FB and FF network obtained by replicating $y$ in the FB network. 
(B) A schematic diagram of the network in (A) for gene regulation. 
(C) The structure of the conjugate FB and FF network obtained by replicating $x$ in the FB network. 
(D) A schematic diagram of the network in (C) for gene regulation.}
\label{fig4}       
\end{figure}

As a more promising strategy for the measurement of the loop gains, we propose a conjugate FB and FF network that is an extension of the dual reporter system for the estimation of the intrinsic and the extrinsic components.
In the conjugate network, we couple the original FB network with a replica that is opened as in \fgref{fig4} (A).
$x$ and $y$ are the same as the original FB network. 
The replica, $y'$, is regulated by $x$ as $y$ is but does not regulate $x$ back.
Thus, $x$ and the replica, $y'$, form an FF network.
If $x$ and $y$ are regulatory proteins as in \fgref{fig2} (B), the replica $y'$ can be engineered by duplicating the gene $y$ with its promoter site, and by modifying the coding region of the replica so that $y'$ looses the affinity for binding to the regulatory region of $x$ as in \fgref{fig4} (B). 
This engineering is much easier than that required for designing the opened network.

Next, we show how to use this conjugate network to measure the loop gains.
For  this network, $K$ and $D$ in \eqnref{eq:LE} become
\begin{align}
K&=\begin{pmatrix}
-d_{x} & k_{xy} & 0 \\
k_{yx} & -d_{y} & 0\\
k_{y'x} & 0  & -d_{y'}
\end{pmatrix},\qquad  
D=\begin{pmatrix}
\bar{a}_{x} & 0 &0\\
0 & \bar{a}_{y} &0 \\
0 &0& \bar{a}_{y'}
\end{pmatrix}.
\end{align}
Because the replica $y'$ affects neither $x$ nor $y$, the fluctuation of $x$ and that of $y$ are the same as those of the FB network in  \eqnref{eq:FDFB}.
The variance of the replica $y'$ can be decomposed as
\begin{align}
\begin{split}
\sigma_{y'}^{2}=&\overbrace{G_{y'y'}\bar{y}'}^{(II)} + \frac{(1-L_{y})+(1+L_{y'})}{(1-L_{x}-L_{y})(1-L_{y'})}\overbrace{G_{y'x}G_{xx}\bar{x}}^{(III)} \\
&+ \frac{1}{(1-L_{x}-L_{y})(1-L_{y'})}\underbrace{(G_{y'y}+G_{y'xy})G_{yy}\bar{y}}_{(V)},
\end{split}
\end{align}
where 
\begin{align}
 L_{y'}=\frac{L_{y}}{2}, \quad G_{y'y'}=G_{yy},\quad G_{y'x}=G_{yx},\quad G_{y'y}=G_{yy'},\quad G_{y'xy}=G_{yx}G_{xy}.
\end{align}
Rearrangement of this equation leads to 
\begin{align}
\sigma_{y'}^{2}&=\sigma_{y}^{2}-\frac{L_{y}}{1-L_{y}/2}G_{yy}\bar{y}.
\end{align}
In addition, we have the following expression for the covariance between $y$ and $y'$ as
\begin{align}
\begin{split}
\sigma_{y,y'}&=\frac{1}{(1-L_{x}-L_{y})}G_{yx}G_{xx}\bar{x}+\frac{(1+L_{x})L_{y'}}{(1-L_{x}-L_{y})(1-L_{y'})}G_{yy}\bar{y},\\
&=\sigma_{y}^{2}-\frac{1}{1-L_{y}/2}G_{yy}\bar{y}.
\end{split}
\end{align}
A similar result can be obtained by replicating $x$ as in \fgref{fig4} (C) and (D).

By using these relations, we have 
\begin{align}
\begin{split}
L_{x}&=\overbrace{\frac{\sigma_{x'}^{2}-\sigma_{x}^{2}}{\sigma_{x,x'}-\sigma_{x}^{2}}}^{(a)}=\overbrace{\frac{1}{1/2+G_{xx}/(F_{x}-F_{x'})}}^{(b)}=\overbrace{2\left(1-\frac{G_{xx}}{F_{x}-F_{x,x'}}\right)}^{(c)},\\
L_{y}&=\underbrace{\frac{\sigma_{y'}^{2}-\sigma_{y}^{2}}{\sigma_{y,y'}-\sigma_{y}^{2}}}_{(a)}=\underbrace{\frac{1}{1/2+G_{yy}/(F_{y}-F_{y'})}}_{(b)}=\underbrace{2\left(1-\frac{G_{yy}}{F_{y}-F_{y,y'}}\right)}_{(c)}, \label{eq:LGconj}
\end{split}
\end{align}
where $F_{x}\defeq \sigma_{x}^{2}/\average{x}$, $F_{y}\defeq \sigma_{y}^{2}/\average{y}$, $F_{x,x'}\defeq \sigma_{x,x'}/\average{x}$, and $F_{y,y'}\defeq \sigma_{y,y'}/\average{y}$ are the Fano factors of $x$ and $y$ and normalized covariances.
This result indicate that we have multiple ways, (a), (b), and (c),  to estimate $L_{x}$ and $L_{y}$ from the statistics of the conjugate network.
In addition, we also have a fluctuation relation that holds between the statistics as
\begin{align}
F_{x}+F_{x'}-2 F_{x,x'}=2G_{xx},\qquad F_{y}+F_{y'}-2 F_{y,y'}=2G_{yy},\label{eq:FR}
\end{align}
which is the generalization of \eqnref{eq:DRSstat} for the dual reporter system.


\begin{figure}
\includegraphics[width=\textwidth]{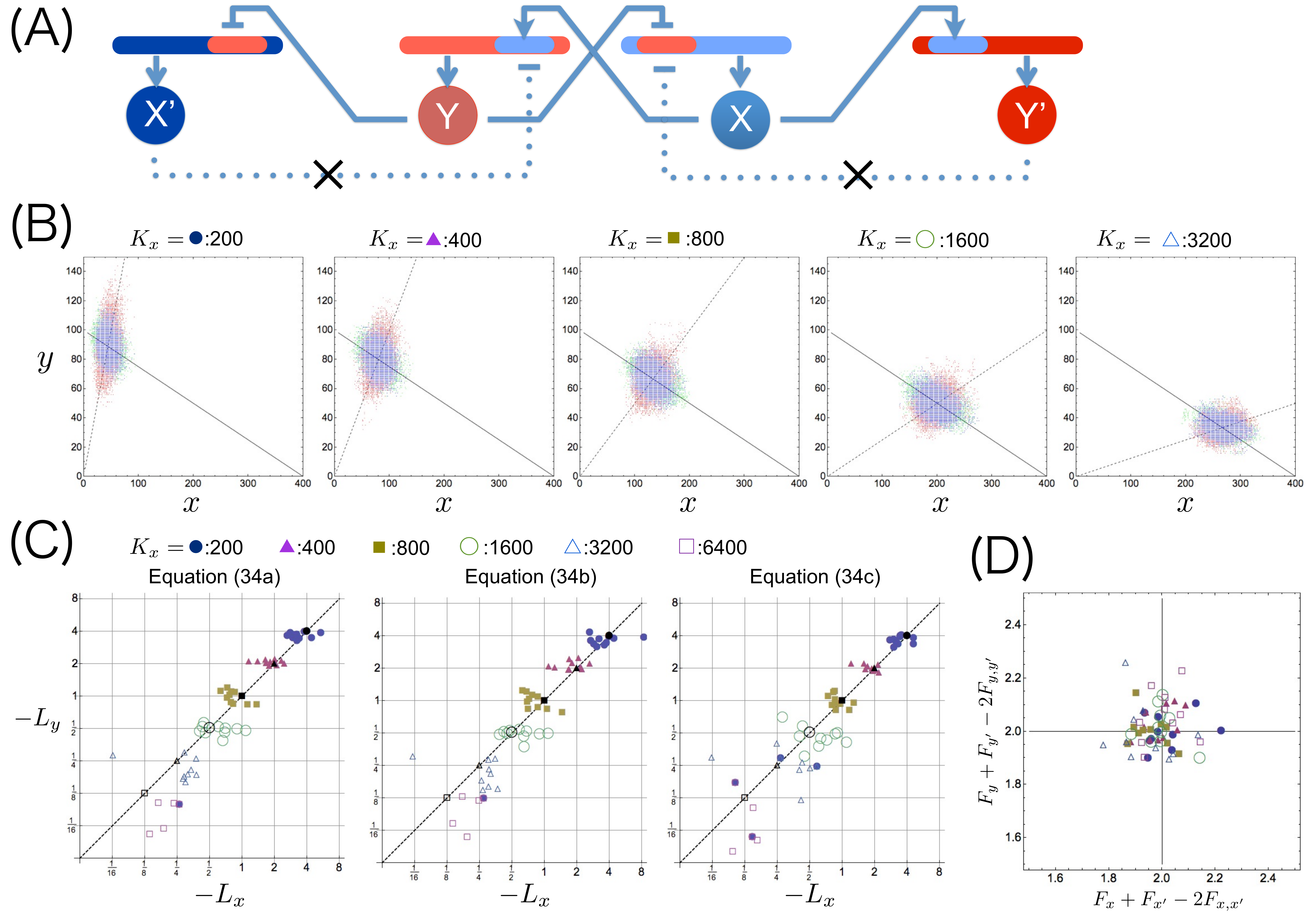}
\caption{(A) A schematic diagram of the conjugate FB and FF network used for the numerical simulation. 
(B) The distributions of $p(x,y)$ , $p(x,y')$, and $p(x',y)$ sampled from the simulation for different parameter values of $K_{x}$. The other parameters are $f_{0}=g_{0}=400$, $K_{y}=100$, and $d_{x}=d_{y}=1$. 
The blue, red, and green dots are the distributions of $p(x,y)$ , $p(x,y')$, and $p(x',y)$, respectively.
Solid and dashed lines are the nullclines of \eqnref{eq:DE} defined as $\dd x/\dt=0$ and $\dd y/\dt=0$ .
(C) $(L_{x}, L_{y})$ estimated with the relations, (a), (b), and (c), in \eqnref{eq:LGconj} from the numerical simulation. 
The parameter values used were the same as in (B).
For each estimation, the means and the covariances required in \eqnref{eq:LGconj} were calculated from $9 \times 10^{4}$ samplings. 
For each parameter value, we calculated the estimates 10 times to see their variation.
Black markers show the analytically obtained values of  $(L_{x}, L_{y})$ for each parameter value.
(D) A plot of the estimates of $F_{x}+F_{x'}-2 F_{x,x'}$ and $F_{y}+F_{y'}-2 F_{y,y'}$ derived in \eqnref{eq:FR}. 
The parameter values used were the same as those in (C). 
}
\label{fig5}       
\end{figure}

\subsection{Verification of the relations by numerical simulation}
We verify \eqnref{eq:LGconj} and  \eqnref{eq:FR} by using numerical simulation.
For the simulation, we use a conjugate network of gene regulation in which the replicas of both $x$ and $y$ are involved to  measure $L_{x}$ and $L_{y}$, simultaneously as in \fgref{fig5}(A). 
For the variable $\Vc{n}=(x,y,x',y')^{\Transpose}$, the stoichiometric matrix and the propensity function are
\begin{align}
S= \begin{pmatrix}
1 & -1 & 0 & 0 & 0 & 0 &0 &0\\
0 & 0 & 1 & -1 & 0 & 0  &0 &0\\
0 & 0 & 0 & 0  & 1 & -1 &0 &0 \\
0&0 &0 & 0 & 0 & 0  & 1 & -1
\end{pmatrix},\quad
a(\Vc{x})= \begin{pmatrix}
f(y) \\ -d_{x} x \\ g(x) \\ -d_{y} y \\ f(y) \\ -d_{x} x' \\g(x) \\ -d_{y} y'  
\end{pmatrix}.
\end{align}
The simulation is conducted by the Gillespie's next reaction algorithm\cite{Gillespie:1976tq}.
First, we test linear negative feedback regulation defined by $f(y)\defeq \max\left[f_{0}[1-\frac{y}{K_{y}}], 0\right]$, and $g(x)\defeq \max\left[g_{0}\frac{x}{K_{x}}, 0\right]$ under which the LNA holds exactly as long as its trajectory has sufficiently small probability to reach the boundaries of $x=0$ or $y=0$.
In \fgref{fig5}(B), the distributions of $p(x,y)$, $p(x,y')$, and $p(x',y)$ are plotted for different parameter values of $K_{x}$.
The FB is strong for small $K_{x}$ whereas it is weak for large $K_{x}$.
$(L_{x}, L_{y})$ estimated by \eqnref{eq:LGconj} for the parameter values are plotted in \fgref{fig5}(C).
The three estimators, (a), (b), and (c) in \eqnref{eq:LGconj}, are used for comparison. 
For this simulation, 
all the estimators work well, but they have slightly larger variability in $L_{x}$ compared with that in $L_{y}$ for large values of $|L_{x}|$ and $|L_{y}|$.
In addition, when $|L_{x}|$ and $|L_{y}|$ are very small, i.e. much less than $1$, 
the estimators show relatively larger variability and bias, suggesting that the estimation of very weak FB efficiency requires more sampling. 
For the same parameter values, we also check \eqnref{eq:FR} in \fgref{fig5}(D).
Both $F_{x}+F_{x'}-2 F_{x,x'}$ and $F_{y}+F_{y'}-2 F_{y,y'}$ localize near $2$ irrespective of the parameter values.
As \fgref{fig5}(C) and (D) demonstrate, the estimators obtained from the simulations agree with the analytical values of  $(L_{x}, L_{y})$, and the fluctuation relation also holds very robustly.

\begin{figure}
\includegraphics[width=\textwidth]{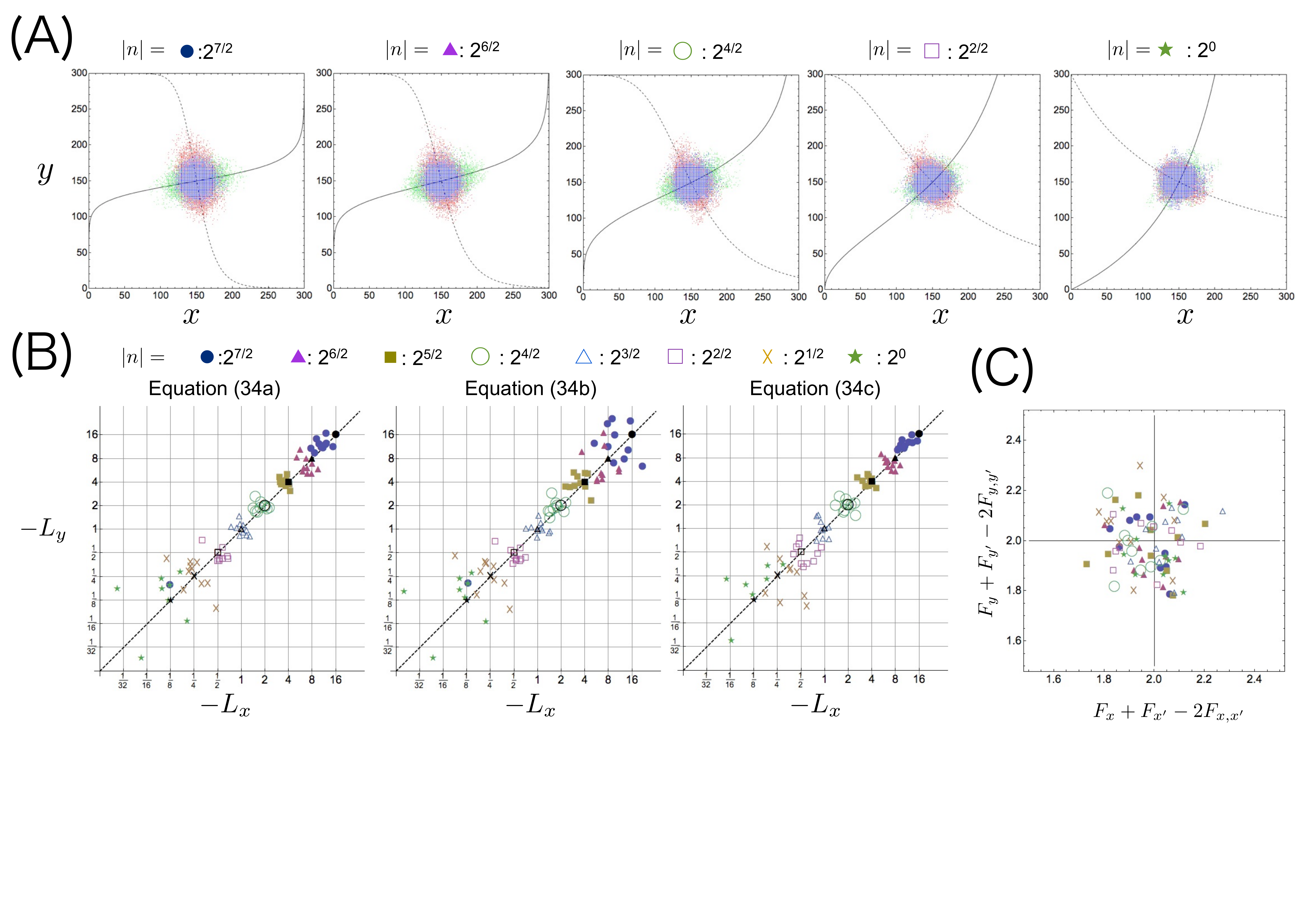}
\vspace{-2cm}
\caption{
(A) The distributions of $p(x,y)$ , $p(x,y')$, and $p(x',y)$ sampled from the simulation for different parameter values of $n_{x}$ and $n_{y}$ where $n_{x}=n_{y}=|n|$.
The other parameters are $f_{0}=g_{0}=300$, $K_{x}=K_{y}=150$, and $d_{x}=d_{y}=1$.
The blue, red, and green dots are the distributions of $p(x,y)$ , $p(x,y')$, and $p(x',y)$, respectively.
Solid and dashed curves are the nullclines of \eqnref{eq:DE} defined as $\dd x/\dt=0$ and $\dd y/\dt=0$ .
(B) $(L_{x}, L_{y})$ estimated with the relations, (a), (b), and (c), in \eqnref{eq:LGconj} from the numerical simulation. 
The same parameter values were used as in (A).
For each estimation,  the means and the covariances required in \eqnref{eq:LGconj} were calculated from $9 \times 10^{4}$ samplings. 
For each parameter value, we calculated the estimates 10 times to see their variation.
Black markers show the analytically obtained values of  $(L_{x}, L_{y})$ for each parameter value.
(C) A plot of the estimates of $F_{x}+F_{x'}-2 F_{x,x'}$ and $F_{y}+F_{y'}-2 F_{y,y'}$ shown in \eqnref{eq:FR}. The parameter values used were the same as those in (B). 
}
\label{fig6}       
\end{figure}
In order to test how nonlinearity affects the estimation,
we also investigated non-linear negative feedback regulation defined by $f(y)\defeq f_{0} \frac{1}{1+(y/K_{y})^{n_{y}}}$, and $g(x)\defeq g_{0} \frac{(x/K_{x})^{n_{x}}}{1+(x/K_{x})^{n_{x}}}$.
We change the Hill coefficients, $n_{x}$ and $n_{y}$, by keeping the fixed point unchanged as in \fgref{fig6}(A).
Compared with the linear case, the variability of the estimators is almost similar even though the feedback regulation is nonlinear (\fgref{fig6}(B)). 
In addition, the estimators show a good agreement with the analytical value except for very large value of $|L_{x}|$ and $|L_{y}|$. 
This suggests that \eqnref{eq:LGconj}  works as good estimators when the trajectories of the system are localized sufficiently near the fixed point as in \fgref{fig6}(A).
For the very large value of $|L_{x}|$ and $|L_{y}|$ where $|n|=2^{7/2}$, 
all the estimators are slightly biased towards smaller values, and (b) in \eqnref{eq:LGconj} has larger variance than the others.
In addition, similarly to the linear case, the estimators require larger sampling when $|L_{x}|$ and $|L_{y}|$ are much less than $1$.
Even with the nonlinear regulation, the fluctuation relation holds robustly as shown in \fgref{fig6}(C), which is consistent with the good agreement of the estimators of $L_{x}$ and $L_{y}$ with their analytical values as in \fgref{fig6}(B).

All these results indicate that the estimators obtained in \eqnref{eq:LGconj} can be used to estimate the FB efficiency as long as the efficiency is moderate and the trajectories of the system are localized near the fixed point.

\section{Discussion}\label{sec:Dis}
In this work, we extended the fluctuation decomposition formula obtained for the FF network to the FB network. 
In this extension, the FB loop gains are naturally derived as the measure to quantify the efficiency of the FB.
By considering the opened FF network obtained by opening the loop of the FB network, the relation between the loop gains and the fluctuation propagation in the FF network was clarified.  
In addition, we proposed the conjugate FB and FF network as a methodology to quantify the loop gains by showing that the loop gains are estimated only from the statistics of the conjugate network. 
By using numerical simulation, we demonstrated that the loop gains can actually be estimated by the conjugate network while we need more investigation on the bias and variance of the estimators.
Furthermore, the fluctuation relation that holds in the conjugate network was also verified. 
We think that our work gives a theoretical basis for the conjugate network as a scheme for experimental estimation of the FB loop gains.

As for the further problems on the efficiency of FB and its quantification, the generality of the FB loop gain should be clarified.
In the case of the FF network, the fluctuation decomposition proposed initially by using the LNA was generalized as the variance decomposition formula with respect to the conditioning of the history of the upstream fluctuation\cite{Hilfinger:2011ed,Bowsher:2012kb}.
In the case of the FB network, similarly, the decomposition and the FB loop gains were obtained and defined via the LNA.
However, its generalization is not straightforward because the contribution of the interlocked components cannot be dissected in the FB network because of the circulative flow of the fluctuation.
The previous work on the relation between the fluctuation decomposition and the information transfer for the FF network\cite{Bowsher:2012kb} suggests that the FB loop gains can also be interpreted in terms of the information transfer. 
As a candidate for such information measure, we illustrate a connection of the conjugate FB and FF network with the directed information and Kramer's causal conditioning.
Let us consider the joint probability of the histories of $x(t)$ and $y(t)$, $\ProbP[\mathcal{X}(t),\mathcal{Y}(t)]$. 
From the definition of the conditional probability, we can decompose this joint probability as
\begin{align}
\ProbP[\mathcal{X}(t),\mathcal{Y}(t)]=\ProbP[\mathcal{X}(t)|\mathcal{Y}(t)]\ProbP[\mathcal{Y}(t)]=\ProbP[\mathcal{Y}(t)|\mathcal{X}(t)]\ProbP[\mathcal{X}(t)].
\end{align}
However, when $x(t)$ and $y(t)$ are causally interacting, we can decompose the joint probability differently as 
\begin{align}
\ProbP[\mathcal{X}(t),\mathcal{Y}(t)]&=\left[ \prod_{i=1}^{t}\ProbP[x_{i}|\mathcal{X}_{i-1},\mathcal{Y}_{i-1}]\right]\left[ \prod_{i=1}^{t}\ProbP[y_{i}|\mathcal{X}_{i},\mathcal{Y}_{i-1}]\right]\\
& = \ProbP_{y||x}[\mathcal{Y}(t)||\mathcal{X}(t)]\times \ProbP_{x||y}[\mathcal{X}(t)||\mathcal{Y}(t-1)],
\end{align}
where $\ProbP_{y||x}[\mathcal{Y}(t)||\mathcal{X}(t)]$ and $\ProbP_{x||y}[\mathcal{X}(t)||\mathcal{Y}(t-1)]$ are the Kramer's causal conditional probability\cite{Permuter:2011jr}.  
If no FB exists from $y$ back to $x$, then this decomposition is reduced to
\begin{align}
\ProbP[\mathcal{X}(t),\mathcal{Y}(t)]
& = \ProbP_{y||x}[\mathcal{Y}(t)||\mathcal{X}(t)]\times \ProbP[\mathcal{X}(t)].
\end{align}
The directed information from $y$ to $x$ is defined as
\begin{align}
\mathbb{I}[\mathcal{Y}(t)\to \mathcal{X}(t)] \defeq \average{\ln\frac{\ProbP[\mathcal{X}(t), \mathcal{Y}(t)]}{\ProbP_{y||x}[\mathcal{Y}(t)||\mathcal{X}(t)]\ProbP[\mathcal{X}(t)]}}_{\ProbP[\mathcal{X}(t), \mathcal{Y}(t)]},
\end{align}
where the joint probability of $\mathcal{X}(t)$ and $\mathcal{Y}$(t) is compared with the distribution, $\ProbP_{y||x}[\mathcal{Y}(t)||\mathcal{X}(t)]\times \ProbP[\mathcal{X}(t)]$\cite{Permuter:2011jr}.
Thus, $\mathbb{I}[\mathcal{Y}(t)\to \mathcal{X}(t)]$ is zero when no FB exists from $y$ to $x$, and measures the directional flow of information from $y$ back to $x$. 
In the conjugate network, the replica $y'$ is driven only by $x$.
Thus, the joint probability between $\mathcal{X}(t)$ and $\mathcal{Y}'(t)$ becomes
\begin{align}
\ProbP[\mathcal{X}(t),\mathcal{Y}'(t)]=\ProbP_{y||x}[\mathcal{Y}'(t)||\mathcal{X}(t)]\ProbP[\mathcal{X}(t)].
\end{align}
Thereby, in principle\footnote{In practice, measuring the joint probability of histories is almost impossible.}, the directed information can be calculated by obtaining the joint distributions, $\ProbP[\mathcal{X}(t),\mathcal{Y}(t)]$ and $\ProbP[\mathcal{X}(t),\mathcal{Y}'(t)]$, of the conjugate network.
This relation of the conjugate network with the directed information strongly suggests that the directed information and the causal decomposition are relevant for the loop gains.
Resolving this problem will lead to more fundamental understanding of the FB in biochemical networks because the directed information is found fundamental in various problems such as the information transmission with FB\cite{Yang:2005cb}, gambling with causal side information\cite{Permuter:2011jr}, population dynamics with environmental sensing\cite{Rivoire:2011fy},  and the information thermodynamics with FB\cite{Sagawa:2012wi}. 
This problem will be addressed in our future work.

\begin{acknowledgements}
We thank Yoshihiro Morishita, Ryota Tomioka, Yoichi Wakamoto, and Yuki Sughiyama for discussion.
This research is supported partially by Platform for Dynamic Approaches to Living System from MEXT, Japan, the Aihara Innovative Mathematical Modelling Project, JSPS through the FIRST Program, CSTP, Japan,  and the JST PRESTO program.
\end{acknowledgements}


\end{document}